\documentclass{article}
\usepackage{emulateapj,epsfig}

\slugcomment{To appear in ApJ Letters, 2000}

\makeatletter

\newenvironment{inlinefigure}{%
\def\@captype{figure}%
\noindent\begin{minipage}{0.999\linewidth}\begin{center}}
{\end{center}\end{minipage}\smallskip}
\makeatother

\begin{document}

\title{Weakly self-interacting dark matter and the structure of dark
halos}

\author{Naoki Yoshida\altaffilmark{1}, Volker
Springel\altaffilmark{1,2}, Simon D.M. White\altaffilmark{1}}
\affil{Max-Planck-Institut f\"{u}r Astrophysik Karl-Schwarzschild-Str.
1, 85748 Garching, Germany}
\and
\author{Giuseppe Tormen\altaffilmark{3}}
\affil{Dipartimento di Astronomia, Universita di Padova,
vicolo dell'Osservatorio 5, 1-35122 Padova, Italy}

\altaffiltext{2}{present address: Harvard-Smithsonian Center for
Astrophysics, 60 Garden Street, Cambridge, MA 02138, USA}

\begin{abstract}
We study the formation of dark halos in a $\Lambda$CDM universe under
the assumption that Cold Dark Matter particles have a finite
cross-section for elastic collisions. We compare evolution when CDM
mean free paths are comparable to halo sizes with the
collisionless and fluid limits. We show that a few collisions
per particle per Hubble time at halo centre can substantially 
affect the central density profile. Cross-sections an order of
magnitude larger produce sufficient relaxation for rich clusters 
to develop core radii in the range 100-200 $h^{-1}$kpc. The structural
evolution of halos is a competition between collisional relaxation
caused by individual particle interactions and violent relaxation 
resulting from the infall and merging processes by which clusters grow.
Although our simulations concentrate on systems of cluster size, we
can scale our results to address the halo structure expected for dwarf
galaxies. We find that collision cross-sections sufficiently large to 
significantly modify the cores of such galaxies produce cluster
cores which are too large and/or too round to be
consistent with observation. Thus the simplest model for
self-interacting
dark matter is unable to improve fits to published dwarf galaxy 
rotation curves without violating other observational constraints. 

\end{abstract}


\section{Introduction}

Recent measurements of structure in the microwave background
radiation (Lange et al. 2000; Hanany et al. 2000), although 
eliminating the ``concordance'' model (Bahcall et al. 1999),
provide strong support for the general theoretical paradigm on which
this model was based. Such Cold Dark Matter (CDM) universes
are in excellent agreement with observed large-scale structure, but 
may be inconsistent with the
observed structure of {\it nonlinear} dark matter dominated systems.
Navarro, Frenk \& White (1997, NFW hereafter) claimed that the 
density profiles of virialized CDM halos 
are reasonably approximated by a ``universal'' form with singular
behavior near it center. More recent simulations with higher
resolution have confirmed this result, suggesting that the central
cusps may be even steeper than the NFW profile
(Moore et al. 1999b; Klypin et al. 1999, see also Jing \& Suto
2000). Such structures appear inconsistent with  published data on 
the rotation curves of dwarf galaxies (Moore 1994; Flores and Primack 
1994) although this inconsistency may reflect limitations of the data
rather than of the theory (van den Bosch et al. 1999; van den Bosch \&
Swaters 2000). There may also be a discrepancy between the
rich substructure seen in simulations of CDM halos
and the relatively small number of satellite galaxies observed
in the Milky Way's halo (Moore et al. 1999a; Klypin et al. 1999).

Spergel \& Steinhardt (2000) suggested that a finite cross-section
for elastic collisions, such that the mean free path of CDM particles 
is short in halo cores but long in their outer parts, might alleviate 
these difficulties. Their proposal has attracted considerable attention.
Ostriker (2000) argued that the massive black holes could grow naturally 
at the centers of galactic spheroids through the accretion of such 
dark matter. Miralda-Escude (2000) pointed out that collisional dark
matter
might produce galaxy clusters which are rounder than observed. 
Mo \& Mao (2000) and Firmani et al. (2000) investigated how
self-interacting dark matter might effect galaxy rotation curves.
Hogan \& Dalcanton (2000) considered how the structural properties
of halos might scale with their mass.
Burkert (2000) and Kochanek \& White (2000) studied how collisional
relaxation would affect the structure of isolated equilibrium halos,
while Moore et al. (2000) and Yoshida et al. (2000) simulated
cluster evolution in a cosmologically realistic context but in the
fluid limit (very short mean free path). In this limit
collisonal dark matter produces more cuspy profiles than collisionless
CDM, and so gives even poorer fits to published rotation curves for
dwarf galaxies.

In this {\it Letter} we continue exploring how collisions
affect the structure of dark halos.
We simulate the formation of a massive halo in a $\Lambda$CDM 
universe assuming scattering cross-sections 
varying over a wide range. The inclusion of the infall and merging
which occur when halos are embedded in their proper cosmological
context leads to core evolution which is considerably more complex than
the expansion followed by collapse seen in the simulations
of Burkert(2000) and Kochaneck \& White(2000).
Cross-sections which would significantly modify the core structure of
dwarf
galaxies produce galaxy cluster cores which are inconsistent with
observation. 

\begin{inlinefigure}
\vspace*{0.2cm}\ \\
\resizebox{8.5cm}{!}{\includegraphics{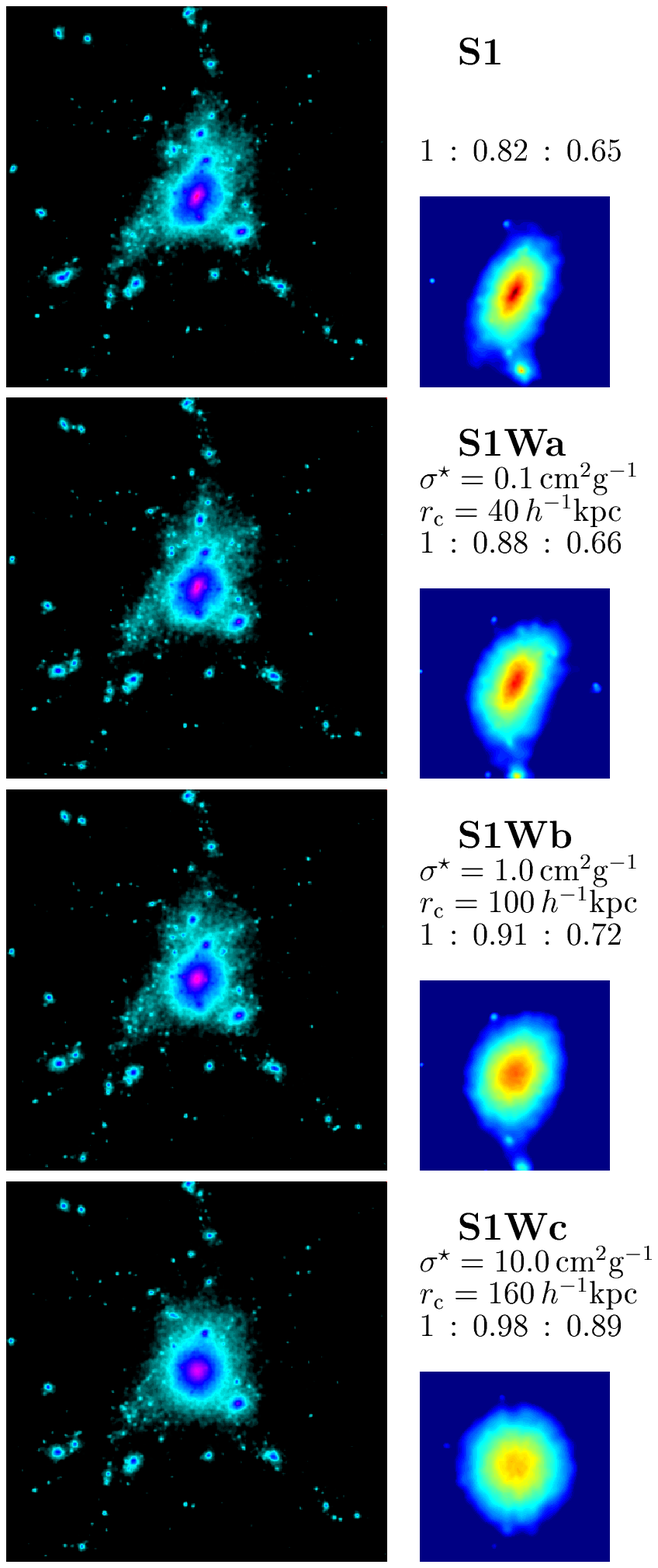}}
\caption{Projected mass distributions in a box
$15h^{-1}$Mpc on a side. The collision cross-sections
per unit mass, core radii, axis ratios for each model 
and small panels showing the central region ($2h^{-1}$Mpc on a side, enlarged) 
in a different color scale are given to the right of the corresponding image. \label{fig1}}
\end{inlinefigure}

\section{THE SIMULATIONS}
Our simulations use the parallel tree code GADGET developed by
Springel (1999, see also Springel, Yoshida \& White 2000). 
We study the same cluster as Yoshida et al. (2000) who 
resimulated the second most massive
object in the $\Lambda$CDM simulation of Kauffmann et al. (1999).
In order to simulate elastic scattering of CDM particles
we adopt the Monte Carlo method introduced by Burkert (2000).
We implement this scheme in the
following manner. At each time step we evaluate the scattering
probability for particle $i$,
\begin{equation}
P = \rho_{i} \sigma^{*} V_{\rm rel} \Delta t,
\label{coll_prob}
\end{equation}
where $\rho_{i}$ is the local density at the particle's position, 
$\sigma^{*}$ is the scattering cross-section per unit mass,
$V_{\rm rel}=|\mbox{\boldmath $v$}_{i}-\mbox{\boldmath $v$}_{\rm ngb}|$
is the
relative velocity between the particle and its nearest neighbour, 
and $\Delta t$ is the time step.
This prescription is similar to Burkert's, but uses the relative
velocity rather than the absolute velocity of particle $i$.
Kochaneck \& White (2000) use a similar scheme but estimate the 
scattering rate more accurately by looping over a certain number of 
neighbours. However, the larger smoothing involved in such a
procedure can itself introduce difficulties in regions with 
significant velocity gradients (Meiburg 1986), and so we
prefer our simpler scheme which should be unbiased even if somewhat
noisier. We choose timesteps small enough to ensure that 
a particle travels only a minor fraction of its
mean free path within $\Delta t$. We assume each collision to be
elastic, 
of hard-sphere type, and to have a cross-section independent of 
velocity. Scattering is assumed isotropic
in the center-of-mass frame, so that relative velocities are randomly
reoriented in each collision. We carry out simulations for three
values of $\sigma^*$ differing by factors of ten.

Most of our simulations employ 0.5$\times 10^{6}$ particles 
in the high resolution region, with a mass per particle 
$m_{\rm p}=0.68\times 10^{10}h^{-1}M_{\odot}$. The gravitational
softening
length is set to 20$h^{-1}$kpc, which is $\sim$1.4\% of the virial
radius of the final cluster. We ran one simulation with 5 times better
mass resolution and 7 times better spatial resolution to check for
numerical convergence. All of our resimulations start from the same
initial conditions. The background cosmology is flat with matter 
density $\Omega_{\rm m}=0.3$,
cosmological constant $\Omega_{\Lambda}=0.7$ and expansion rate
$H_{0}=70$ km$^{-1}$Mpc$^{-1}$.  It has a CDM power spectrum normalised
so that $\sigma_{8}=0.9$. The virial mass of the final cluster is
$M_{200}=7.4\times 10^{14}h^{-1}M_\odot$, determined as the mass within
the radius $R_{200}= 1.46 h^{-1}$Mpc where the enclosed mean
overdensity is 200 times the critical value. 

\section{RESULTS}
The large-scale matter distribution in all our simulations 
looks very similar. Because we start from identical 
initial conditions, the particle distributions 
differ only in regions where collisions are important.
Figure 1 shows that the final cluster is more nearly 
spherical and has a larger core radius for larger
collision cross-section. The quoted axial ratios are determined 
from the inertia tensors of the 
matter at densities exceeding 100 times the critical value. 
Miralda-Escude (2000) argues that the ellipticity of
cluster cores, as inferred from gravitational lensing observations,
can be used to limit the interaction cross-section. 
Among our final clusters, S1W-b and S1W-c are severely constrained 
by the limits he quotes. 

In Figure 2 we show density profiles for all of our simulations.
Also plotted in the bottom panel is the mean collision number
per particle. (We counted collisions for each
particle throughout the simulation.) Figure 2 clearly shows the
presence of a core whose extent depends on the cross-section. For 
our intermediate cross-section case (S1W-b), we also carried out a
higher
resolution simulation. The two density profiles agree very well
(see Figure 2) showing that our simulations have converged 
numerically on scales larger than the gravitational softening length. 
The mean collision count at cluster center is 3
for S1W-a, 8 for S1W-b, and 35 for S1W-c. Thus only a few
collisions per particle in a Hubble time suffice to affect
the central density profile, and about 10 collisions per particle
result in a core with $r_{\rm c}$ $\ge$ 100$h^{-1}$kpc.
(We define the core radius as the point
where the density profile becomes steeper than $\propto r^{-1}$.
Core radii by this definition are given on the right of Figure 1.)

\begin{inlinefigure}
\vspace*{0.2cm}\ \\
\resizebox{8cm}{!}{\includegraphics{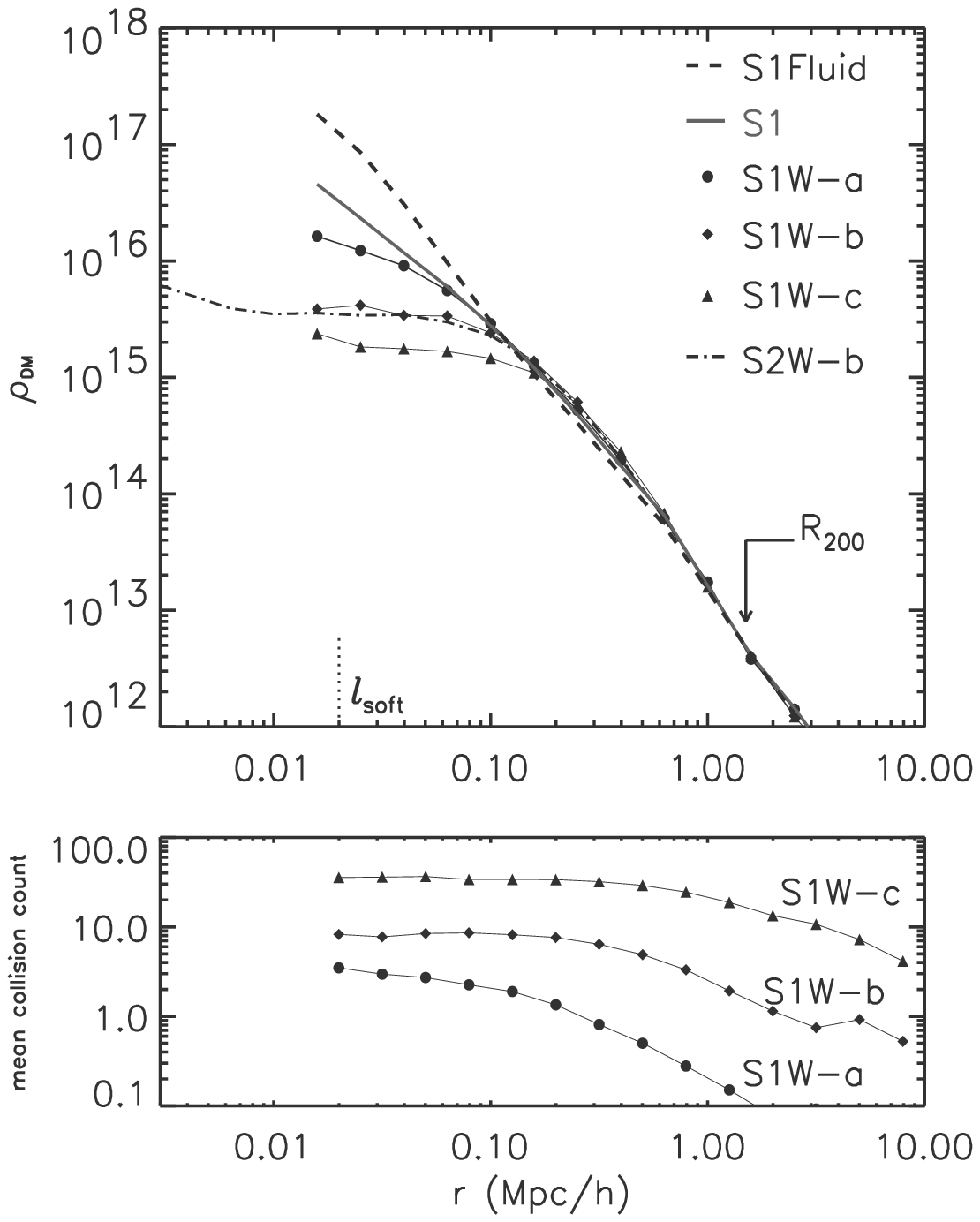}}
\caption{Density profiles (top) and mean collision counts
per particle (bottom). The vertical dotted line in the top panel
indicates the gravitational softening length of our S1 simulations. The virial
radius $R_{200}$ of the final cluster is shown as an arrow. 
The ``fluid" dark matter case from Yoshida et al. (2000) is 
plotted as the dashed line, while the
dash-dotted line represents our higher resolution simulation
of the medium cross-section case (S2W-b). \label{fig2}}
\end{inlinefigure}

Unlike an isolated system, our cluster grows through successive mergers.
Thus the material in its central region is a mixture of the material 
from a number of its progenitors. Figure 3, a time sequence of density
profiles for our S1W-c simulation, shows clearly how merger events 
interrupt the core evolution and produce low density cores. 
We let this simulation run beyond the present time to a=1.72,
where $a$ is the expansion parameter normalised to its present value.
During the time interval plotted, the cluster experiences major mergers
at $a\sim 0.75$ and $a\sim 1.4$. Each of these events is associated
with an increase of the core radius. Subsequent relaxation
causes the core to shrink again and the central density to rise. For
this
relatively large cross-section, relaxation-driven core expansion does
not
occur within the time interval shown. In contrast to this complex
behavior, the virial mass of the cluster grows quite smoothly, 
approximately doubling between $a=0.73$ and $a=1.72$. Clearly, the 
core radius of the cluster at $a=1$ results from an interplay between
collisional relaxation driven by particle collisions and violent
relaxation caused by mergers.

\begin{inlinefigure}
\vspace*{0.3cm}\ \\
\resizebox{8cm}{!}{\includegraphics{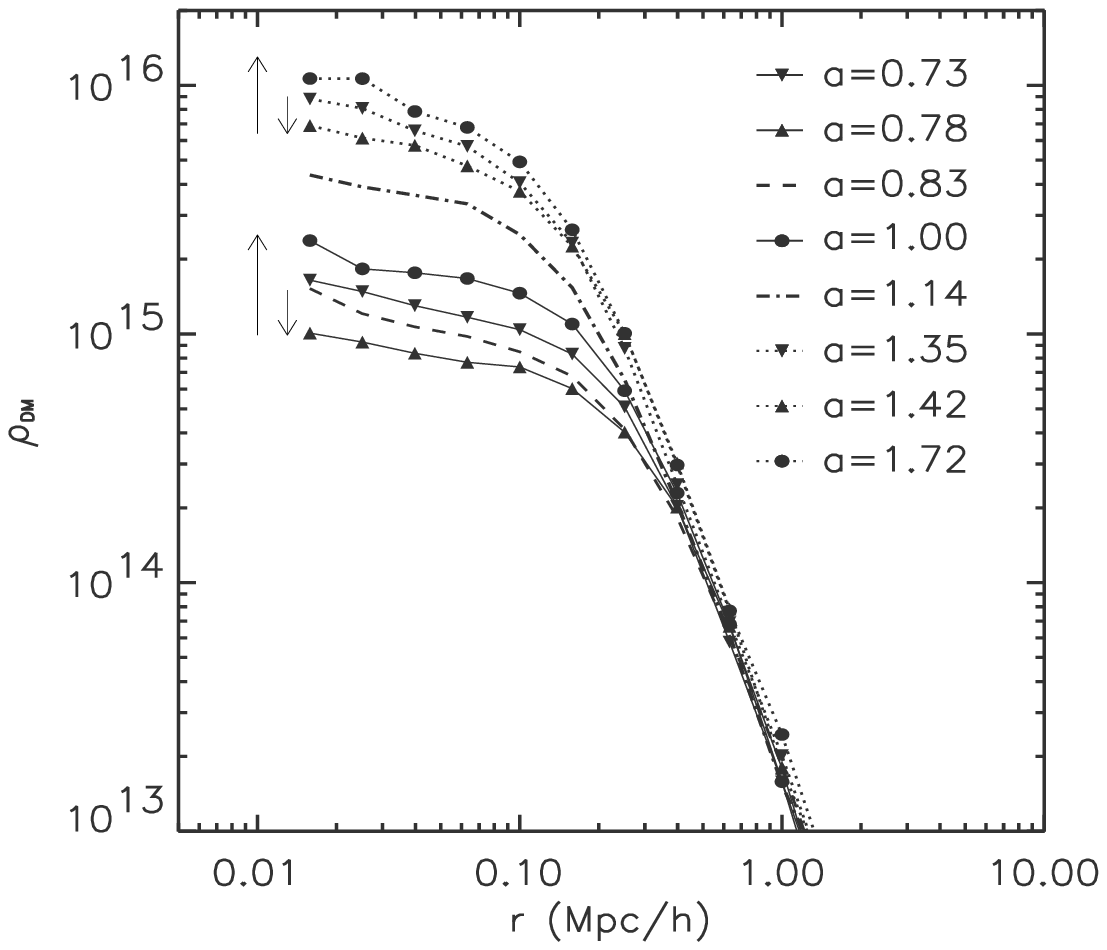}}
\caption{We plot the time evolution of the density profile
in S1W-c (the largest cross-section case).
Time is indicated by the expansion parameter $a$.
The series starts at $a=0.73$ just before the final merger occurs.
After the merger the core settles down again by $a=0.78$ 
and then enters a core-collapse phase which is interrupted by another
major merger at $a=1.4$. \label{fig3}}
\end{inlinefigure}

\begin{inlinefigure}
\vspace*{0.3cm}\ \\
\resizebox{8cm}{!}{\includegraphics{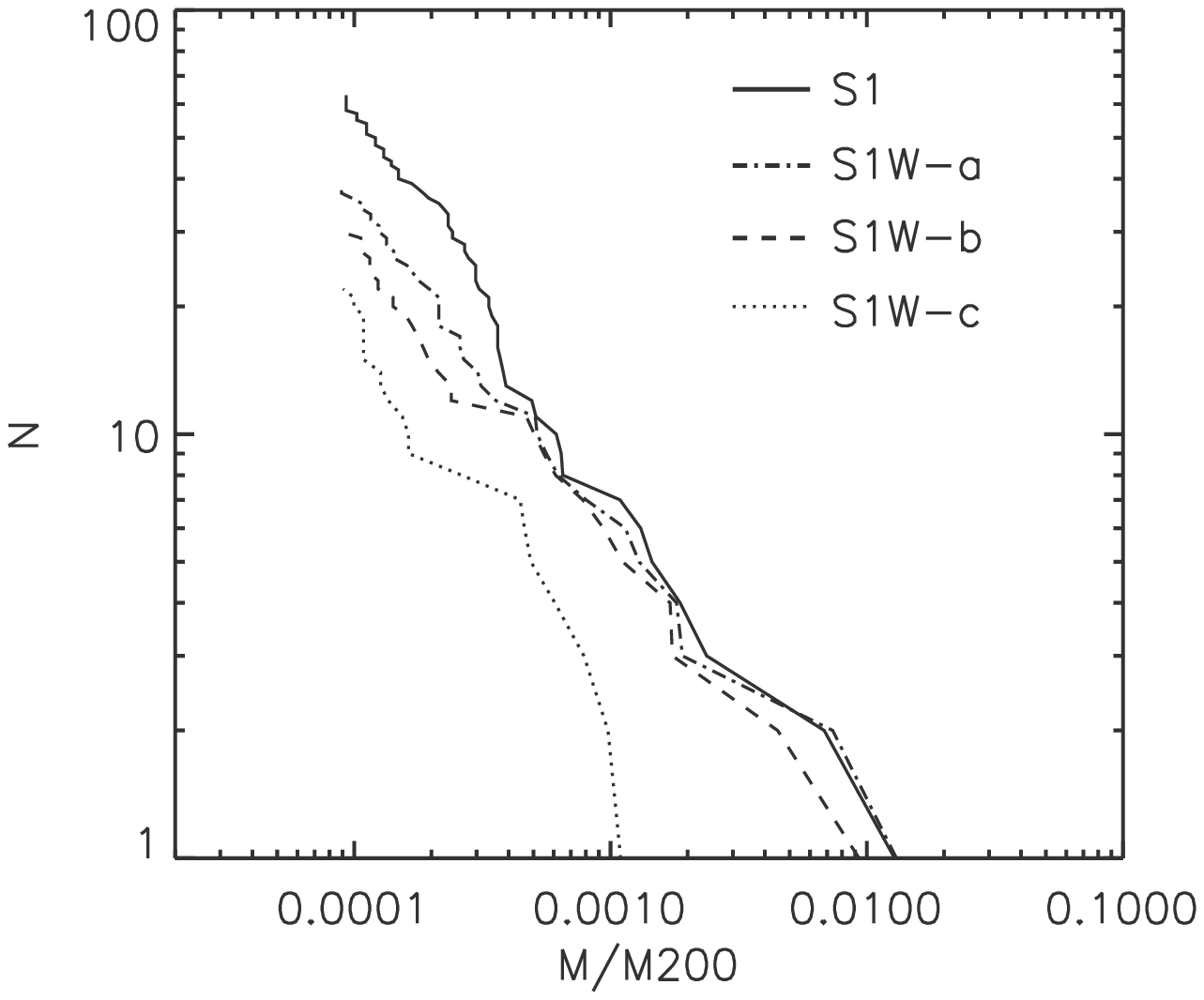}}
\figcaption[fig4.ps]{The total number of subhalos within $R_{200}$ is
plotted as a function of the lower limit to their mass in units of $M_{200}$.
Results are plotted for halos containing 10 or more particles.\label{fig4}}
\end{inlinefigure}
 
In Figure 4 we compare the amount of substructure within $R_{200}$
in our various simulations.  We use the SUBFIND algorithm by
Springel (1999) to identify subhalos in the final cluster.
This identifies gravitationally self-bound sets of particles that are
at higher density than the smooth background of cluster material.
Local density is defined at each particle's position in a SPH fashion.
Using this procedure we find that 3.7\%, 3.6\%, 2.5\%, and 0.7\%
of the cluster mass is included in subhalos in S1, S1W-a, S1W-b and
S1W-c respectively. Although low mass substructures are somewhat 
less abundant for larger cross-sections, massive subhalos are {\it
not} substantially disrupted in S1W-a and S1W-b. Many of the massive
subhalos are $\sim 1h^{-1}$Mpc from the cluster center, where
particle collisions are rare in these models (see the bottom panel of
Figure 2). Hence ``dark matter evaporation" (Spergel \&
Steinhardt 2000) is ineffective for them.
On the other hand, the massive subhalos in S1W-c are totally
disrupted. Infalling halos are rapidly stripped by collisions with
``diffuse'' cluster dark matter in this case.

\section{Summary and Discussion}

Our simulations of cluster formation in a $\Lambda$CDM
universe made of self-interacting dark matter show that 
collision rates exceeding one or two per particle per Hubble time
at cluster center are sufficient to produce a constant density
core. Observations of strong lensing by clusters require their 
cores to be dense and small. Thus to fit
the cluster Cl0024+1654 Tyson et al (1998) needed a core radius
of $35h^{-1}$~kpc and a central surface density of $7900 
h$~M$_\odot/$pc$^2$. Recent HST observations of an unbiased sample of
X-ray luminous clusters at $z\sim 0.2$ find thin giant arcs at
similar radii (10 to 25 arcsec) in almost all of them (J.-P. Kneib,
private communication) showing Cl0024+1654 to be quite typical. We 
find core radii of this order in our
simulations for a cross-section of 0.1 cm$^2$/g (see Figure 1).

Predicted collision rates in dwarf galaxy cores are much
smaller than in clusters for the core radii usually inferred 
from rotation curve data. Thus for the archetypal example DDO154,
Carignan and Beaulieu (1989) give as best fit parameters
a core radius $r_c = 3$~kpc and a central surface density
$\Sigma_o = 141$~ M$_\odot/$pc$^2$. (Note the lack of $h$ dependence 
because the distance is measured directly.) The central
collision rate scales as $\Sigma_o^{1.5}/r_c^{0.5}$, so collisions are
inferred to be 60 times less frequent in DDO154 than in Cl0024+1654
(for $h=0.7$).
Since particles in the cluster core have no more than a couple
of collisions in a Hubble time (see Figure 2) it is difficult to see
how collisions could produce the large apparent core in
DDO154. This agrees with the recent results of Dav\'{e} et al.
(2000) who concluded from their own simulations that cross-sections 
of order 5 cm$^2$/g are needed to produce 
good agreement with the apparent cores of dwarf galaxies; they found
at best marginal consistency for 0.5 cm$^2$/g, a value which is still
5 times the upper limit we find to be consistent with cluster data.

A possible solution might seem to be a cross-section
about two orders of magnitude larger than in S1W-c. Dwarf
galaxy haloes would then look similar to a scaled version of
S1W-c, with a core radius of about $4h^{-1}$kpc for the parameters
considered above, while rich clusters would be highly collisional and
might have profiles approaching that in our ``fluid'' 
simulation. Our earlier work confirmed, however, that such clusters
would be almost spherical; such large cross-sections can thus
be excluded following Miralda-Escude's (2000) argument.

A different resolution might be to introduce 
an interaction law which implies an energy dependent cross-section
such that scattering is less 
effective in high velocity encounters. This would reduce the
difference between cluster and dwarf galaxy halos. 
This idea requires a more detailed physical model for the
dark matter, and we do not pursue it further here. We
note that $\sigma^*\propto V^{-1}$ is required to make the
collision rate at $r_{\rm s}$ approximately independent of halo mass.
Another loophole might be for the cores of clusters to contain large
amounts of baryonic dark matter, perhaps deposited by cooling flows.
The cooling rates inferred from X-ray data appear too small for 
this to be viable (see, e.g. Peres et al. 1998). 

In summary our results suggest that collisional dark matter 
cannot produce core radii in dwarf galaxy halos as large as
those inferred from rotation curve observations without
simultaneously producing cluster cores which are too large and too
round to be consistent with gravitational lensing data.

\acknowledgments
The authors thank Tom Abel, Jerry Ostriker and Paul Steinhardt
for fruitful discussions.
The hospitality of the Institute for Theoretical Physics,
UCSB, where this work was initiated, is also acknowledged.

\end{document}